\documentclass[aps,prd,nofootinbib]{revtex4}
\usepackage[colorlinks=true, pdfstartview=FitV, linkcolor=blue, citecolor=red, urlcolor=magenta]{hyperref}
\usepackage{graphicx}
\usepackage{latexsym}
\usepackage{amsmath}
\usepackage{amsfonts}
\usepackage{amssymb}
\usepackage{verbatim}
\usepackage{braket}
\usepackage{extarrows}

\newcommand{\be}{\begin{equation}}
\newcommand{\ee}{\end{equation}}
\newcommand{\bea}{\begin{eqnarray}}
\newcommand{\eea}{\end{eqnarray}}

\newcommand{\ben}{\begin{eqnarray}}
\newcommand{\een}{\end{eqnarray}}

\begin{document}

\title{Thermal Casimir effect for the scalar field in flat spacetime under a helix boundary condition}

\author{$^{1}$Giulia Aleixo}
\email{giulia.aleixo@academico.ufpb.br}

\author{$^{1}$Herondy F. Santana Mota}
\email{hmota@fisica.ufpb.br}

\affiliation{$^{1}$Departamento de F\' isica, Universidade Federal da Para\' iba,\\  Caixa Postal 5008, Jo\~ ao Pessoa, Para\' iba, Brazil.}


\begin{abstract}
In this work we consider the generalized zeta function method to obtain temperature corrections to the vacuum (Casimir) energy density, at zero temperature, associated with quantum vacuum fluctuations of a scalar field subjected to a helix boundary condition and whose modes propagate in (3+1)-dimensional Euclidean spacetime. We find closed and analytical expressions for both the two-point heat kernel function and free energy density in the massive and massless scalar field cases. In particular, for the massless scalar field case, we also calculate the thermodynamics quantities internal energy density and entropy density, with their corresponding high- and low-temperature limits. We show that the temperature correction term in the free energy density must suffer a finite renormalization, by subtracting the scalar thermal blackbody radiation contribution, in order to provide the correct classical limit at high temperatures. We check that, at low temperature, the entropy density vanishes as the temperature goes to zero, in accordance with the third law of thermodynamics. We also point out that, at low temperatures, the dominant term in the free energy and internal energy densities is the vacuum energy density at zero temperature. Finally, we also show that the pressure obeys an equation of state.
\end{abstract}
 \maketitle


\section{Introduction}
\label{intro}
%
 The Casimir Effect was predicted for the first time back in 1948 by Hendrik Casimir \cite{Casimir1948dh}. Ten years later, in 1958, an experiment conducted by M. Sparnaay \cite{Sparnaay1958} showed compatibility between Casimir's theoretical results and experimental data. However, the accuracy of the experiment at the time was not enough to precisely confirm it. Although Casimir's prediction did not receive much attention in its first years, it began to attract attention even before later accurate experimental verifications conducted at the end of the century by Lamoreaux, Mohideen and Roy \cite{PhysRevLett.81.5475, Lamoreaux1996wh, MohideenRoy1998iz} (see also Refs. \cite{Bressi:2002fr, MOSTEPANENKO2000, PhysRevA.78.020101, PhysRevA.81.052115}). The study of this phenomenon have drawn increasing attention over time, not only for its theoretical and mathematical aspects but also for the numerous and pertinent applications given that it is a quantum effect with macroscopic manifestations \cite{BordagMohideenMostepanenko, Mostepanenko:1997sw, bordag2009advances, Milton:2001yy}. 
    
    The Casimir effect arises due to perturbations in the vacuum state fluctuations of a given quantum field, generating a force that could be attractive or repulsive depending on the conditions that cause the vacuum energy to fluctuate. The modifications in the fluctuations of the quantum field can be induced by a series of factors such as the imposition of boundaries conditions, the spacetime topology and dimensionality, the nature of the background field and as we discuss in this work, the temperature \cite{BordagMohideenMostepanenko, Mostepanenko:1997sw, bordag2009advances, Milton:2001yy, Edery:2006td, Edery:2005bx, Edery:2005aj, Edery:2007xf}. 
    
    The analysis of the Casimir forces is conducted by means of powerful regularization and renormalization techniques for the vacuum energy. Among them, the $\zeta$-function provides an elegant approach, which has been extensively developed and discussed for decades \cite{Hawking1977, Elizalde2012zv, Edery:2006td, Edery:2005bx, Edery:2005aj, Edery:2007xf, Lin:2014lva}. Rigorous and comprehensive applications of the $\zeta$-function method can be found in \cite{Elizalde1994book}. In particular, as introduced by J. S. Dowker and R. Critchley \cite{DowkerCritchley1975tf} and reshaped by Hawking \cite{Hawking1977}, a connection can be made between a generalized $\zeta$-function and the determinant of differential operators such as the D'Alambertian and the Laplace-Beltrami operators. In this context, the path integral approach shows to be of great importance since it allows us to obtain a regularized and, consequently, renormalized expression for the energy density from the partition function of the field, in terms of the corresponding operator, providing a way of exploiting temperature contributions.

    As technology advancements set foot into the micro and nanoscopic scales, the Casimir effect becomes a key element to understand and predict numerous phenomena. In biophysics, for instance, as devices approach the single cell size, recent studies indicate that the Casimir and Casimir-like forces may play a significant role on how cells can adhere to one another (despite the negative charge some of them carry) or how it may also affect the protein folding, the wettability of some structures and a number of others cellular interactions \cite{brandonjicRedBloodCells10, geometricpot_wettability76, CasimirUniversalCellMembrane4, Gambassi2009REV, machta113}.
    
    In nanotechnology the development of nanoelectromechanical and microelectromechanical systems brings up the need to evaluate how the Casimir force acts on material components depending on their properties, such as geometry and temperature. In such tiny devices the Casimir forces could cause the mechanical components to collapse and adhere to nearby surfaces which may result in a device malfunction, or if properly engineered it can give the device an `anti-stiction" property. In particular, if one considers the Casimir effect for nanotubes (cylinder structures at the nanoscale), the link between the quantum field and the structure is represented by a quasi-periodic condition where the phase angle mimics the conductivity properties of the nanotubes \cite{QuasiPNanotubes,klecioQuasiPNanotubes}.
    
    It is clear that understanding the role that boundary conditions, material properties, geometry and temperature play on the vacuum state fluctuations is crucial. In this work we seek to calculate temperature contributions for the Casimir energy density, in a flat spacetime, arising from perturbations on the scalar field vacuum as a consequence of the imposition of a helix boundary condition. There are numerous structures in nature with a helix geometry such as the DNA and cell membrane proteins to motivate the investigation of Casimir forces from such a condition. The force arising from the helix boundary condition at zero temperature has been considered in Refs. \cite{QuantumSpringFromCasimir, QuantumSpring, QuantumSpringD+1} where it was found the vacuum force to be linear in the axis of the helix for small values of the ratio between its pitch and its circumference, being in this case very much like the force on a spring that obeys the Hooke's law. As this phenomenon is the consequence of a quantum effect, the structure that causes it has been called by the authors in \cite{QuantumSpringFromCasimir, QuantumSpring, QuantumSpringD+1} as \textit{quantum spring}.
      
 This work is organized as follows: in Sec.II an overview of the generalized zeta function method is given, along with its connection with the thermodynamics quantities, namely, free energy density, entropy density and pressure. In addition, in Sec.III, we consider the scalar field in flat spacetime under the influence of a helix boundary condition. We also calculate the two-point heat kernel function, the local zeta function and all the relevant thermodynamics quantities. We also analyze the high- and low-temperature limits in each case. Finally, in Sec.IV we present our conclusions. In this paper we use natural units $\:\hbar\:=\:c\:=\:1$.
%
\section{Generalized zeta function method}
In this section we intend to provide a somewhat detailed overview about the generalized zeta function method to implement temperature corrections to the vacuum energy, or even to calculate it at zero temperature. The generalized zeta function can be constructed by using the eigenvalues of a known differential operator, like the Laplace-Beltrami operator, very common in physics. Let us then generically consider a four-dimensional operator $\hat{A}_4$, with eigenvalues $\lambda_j$. The generalized zeta function is, thereby, defined as \cite{Hawking1977, Elizalde2012zv, Elizalde1994book}
\begin{eqnarray}
\zeta_{4}(s) = \sum_j\lambda_j^{-s}.
\label{ZF}
\end{eqnarray}
It converges, in four dimensions, for Re$(s) > 2$ and it is regular at $s=0$. Nevertheless, it can also be analytically extended to a function of $s$, with poles at $s=2$ and $s=1$. Note that the spectrum of eigenvalues of $\hat{A}_4$ may not always be discrete \cite{Elizalde1994book}.

Now, we can use the derivative, with respect to $s$, at $s=0$, of the generalized zeta function \eqref{ZF} and write 
\begin{eqnarray}
e^{-\zeta'_{4}(0)} &=& \prod_j\lambda_j\nonumber\\
&=& \text{det}(\hat{A}_4).
\label{detZF}
\end{eqnarray}
This is a particularly useful identity to obtain the partition function of the system later on, as we shall see.

In order for us to make the connection of the zeta function \eqref{ZF} with the partition function we need to make use of the path integral formulation for quantum field theory. In this sense, firstly one should remind that, in statistical mechanics, the partition function is given by \cite{GibbonsHaking1976, bellac1991quantum, Zinn-Justin:572813}

\begin{eqnarray}
Z &=& \text{Tr}\left[ e^{-\beta\hat{H}}\right]\nonumber\\
&=& \int d\Phi \bra{\Phi}e^{-\beta\hat{H}}\ket{\Phi}.
\label{PF}
\end{eqnarray}
where $\Phi$ is a quantum scalar field, $\hat{H}$ is the hamiltonian operator and $\beta=\frac{1}{k_BT}$, with $T$ being the temperature. Note that here, we are only interested in considering scalar field quantum modes. Note also that the integrand is the probability amplitude responsible for taking the system from an inicial state at a time $t_i$ to a final state at time $t_f$, and must be integrated over all scalar field configurations. In the process of explicitly showing the connection of the partition function \eqref{PF} with the path integral formulation we need to introduce a periodicity in time, that is, $\tau = i(t_f - t_i )=\beta$ \cite{GibbonsHaking1976, Hawking1977}. Consequently, the quantum scalar field $\Phi$ must be periodic in the imaginary time $\tau$, i.e.,
 \begin{equation} \begin{split} \label{period}
        \Phi(\tau) = \Phi(\tau + \beta).
    \end{split}\end{equation}
Moreover, the integral in Eq. \eqref{PF} can in fact be put in the form \cite{GibbonsHaking1976, bellac1991quantum, Zinn-Justin:572813}
    \begin{eqnarray}
        Z = \int \mathcal{D}\Phi \: e^{iI(\Phi)},
                      \label{PFPI}
    \end{eqnarray}
which is given in terms of the quadratic Euclidean action
    \begin{eqnarray}
        I(\Phi) = -\frac{1}{2} \int\sqrt{-g} \Phi\hat{A}_4 \Phi \: d^4x.
        \label{EC}
    \end{eqnarray}
From now on we make clear that $\hat{A}_4$ is an elliptic, self-adjoint, second order differential operator and the metric $g_{\mu\nu}$ is Euclidean, with determinant $g$ \cite{Hawking1977, Elizalde2012zv, Elizalde1994book}. The latter is used to construct the operator $\hat{A}_4$. Typically in physics, we associate this operator with the D'Alembertian $\Box_{\text{E}}$ in the four-dimensional Euclidean spacetime, with the imaginary time $\tau$, that is,
  \begin{eqnarray}\label{Box}
       \hat{A}_4 &=& (\Box_{\text{E}} + m^2)\nonumber\\
       &=& \frac{1}{\sqrt{g}}\partial_i\left[\sqrt{g}g^{ij}\partial_j\right] + m^2,
    \end{eqnarray}
where $m$ is the mass of the scalar field and the Euclidean spacetime metric $g^{\mu\nu}$ has signature $(-1,-1,-1,-1)$. One should note at this point that the present approach can be extended to fermion fields. In this case, the operator $\hat{A}_4$ is of first order and the fermion field is anti-periodic in the imaginary time $\tau$ \cite{Elizalde1994book}.

It is possible to solve the gaussian integral in Eq.\eqref{PFPI}-\eqref{EC} by noting that the operator $\hat{A}_4$ obeys the eigenvalue equation
    \begin{eqnarray}
        \hat{A}_4\phi_j = \lambda_j\phi_j,
        \label{EVE}
    \end{eqnarray}
with the complete set of eigenfunctions $\phi_j$ and eigenvalues $\lambda_j$. The former, as usual, in the four-dimensional Euclidean spacetime, can be normalized such that 
    \begin{eqnarray}
        \int\sqrt{g}\phi_j(x)\phi^*_k(x)d^4x = \delta_{jk},
        \label{Orth}
    \end{eqnarray}
where $x=(x^0,x^1,x^2,x^3)$ in the argument of the eigenfunctions represent spacetime coordinates and $d^4x$ the spacetime volume element.

We want now to solve the Gaussian integral in \eqref{PFPI}-\eqref{EC} by expanding the quantum scalar field $\Phi$ in terms of the eigenfunctions $\phi_j$, that is,
  \begin{eqnarray}
       \Phi = \sum_ja_j\phi_j,
        \label{Ex}
    \end{eqnarray}
where $a_j$ are the coefficients of the expansion. Moreover, the element $\mathcal{D}\Phi $ associated with the integration over all field configurations, consequently, can be written as
  \begin{eqnarray}
      \mathcal{D} \Phi = \prod_{j}\mu da_j,
        \label{Ex}
    \end{eqnarray}
with $\mu$ being a constant with dimension of mass, which has been used to normalize the product in \eqref{Ex}. 

 Upon taking into consideration Eqs. \eqref{EVE}-\eqref{Ex}, the path integral in \eqref{PFPI}-\eqref{EC} is found to be
   \begin{eqnarray}
        Z = \left[\text{det}\left(\frac{4}{\pi\mu^2}\hat{A}_4\right)\right]^{-\frac{1}{2}},
                      \label{Zdet}
    \end{eqnarray}
which can be put in the form
   \begin{eqnarray}
        \ln Z = \frac{1}{2}\zeta_4'(0) + \frac{1}{2}\ln\left(\frac{\pi\mu^2}{4}\right)\zeta_4(0).
                      \label{Zlog}
    \end{eqnarray}
This expression is particularly useful once one knows explicitly the eigenvalues of the operator $\hat{A}_4$, which is normally the case in flat spacetime and in some particular curved spacetimes, like for instance, the one describing a closed Einstein universe \cite{PhysRevD.83.104042, Bezerra:2011nc, Bezerra:2017zqq}.

In curved spacetime, where the eigenvalues are not usually known, the convenient approach to be adopted is by using the heat kernel, which provides information about the spacetime where it is defined. Let us then consider, $K(x,x',\eta)$, as being the heat kernel obeying the heat equation
    \begin{eqnarray}
        \frac{\partial}{\partial\eta}K(x,x',\eta) + \hat{A}_4K(x,x',\eta) = 0.
        \label{HE}
    \end{eqnarray}
Note that the operator $\hat{A}_4$ acts on the Euclidean spacetime coordinates $x$, and $\eta$ is a parameter with dimension of time. If one considers the eigenvalue equation \eqref{EVE}, the solution for the heat equation above is clearly given by
    \begin{eqnarray}
       K(x,x',\eta) = \sum_je^{-\lambda_j \eta}\phi_j(x)\phi^{*}_j(x'),
        \label{HK}
    \end{eqnarray}
with the initial condition
    \begin{eqnarray}
       K(x,x',0) = \delta(x-x').
        \label{HKin}
    \end{eqnarray}

Furthermore, upon using the heat kernel expression \eqref{HK}, along with the normalization condition \eqref{Orth}, we are able to show that 
  \begin{equation}\label{traceK}
        \text{Tr}\left[e^{-\eta\hat{A}_4}\right] \equiv \int \sqrt{g}K(x,x, \eta)d^4x = \sum_j e^{-\lambda_j\eta}.
    \end{equation}
We can now relate the trace above with the generalized zeta function \eqref{ZF}, by means of the Mellin transformation \cite{mellin}
  \begin{equation}
        f(s) = \frac{1}{\Gamma(s)}\int_0^\infty \eta^{s-1}F(\eta)d\eta.
    \end{equation}
In the case the function $f(s)$ is the generalized zeta function defined in \eqref{ZF} and $F(\eta)$ is given by \eqref{traceK}, we have
  \begin{equation}\label{GZeta}
        \zeta_4(s) = \frac{1}{\Gamma(s)}\int_0^\infty \eta^{s-1} \text{Tr}\left[e^{-\eta\hat{A}_4}\right] d\eta.
    \end{equation}
This expression is an alternative representation for the generalized zeta function \eqref{ZF}, which is sometimes also known as global zeta function since it does not depende on the spacetime coordinates. It is very useful to calculate the vacuum free energy when the eigenvalues of the operator $\hat{A}_4$, obeying some boundary condition, are explicitly known. Once one knows the zeta function, of course, we can use Eq. \eqref{Zlog} to calculate the partition function and, consequently, the vacuum free energy. 

In the case the eigenvalues of the operator $\hat{A}_4$ are not explicitly known we can define the local zeta function, in terms of the heat kernel $K(x,x,\eta)$, i.e.,
  \begin{equation}\label{localZeta}
        \zeta_4(x,s) = \frac{1}{\Gamma(s)}\int_0^\infty \eta^{s-1}K(x,x,\eta) d\eta.
    \end{equation}
 This is connected with the vacuum free energy density (see Eq. \eqref{freeEnergyDensity}). One should note that, usually in curved spacetime, it is not always possible to calculate the vacuum free energy and in this situation we can only obtain local quantities, as the vacuum free energy density. This is the case of the cosmic string spacetime, see for instance \cite{BezerradeMello2011nv, MotaDispiration}.

On the other hand, in the case it is possible to obtain the vacuum free energy one can, by integrating in the spacetime coordinates, $x$, the local zeta function $\zeta(x,s)$, calculate the global zeta function. This is done by combining Eqs. \eqref{localZeta} into \eqref{GZeta}, that is, 
  \begin{equation}\label{GZeta2}
        \zeta_4(s) =\int\sqrt{g}\zeta_4(x,s)d^4x.
    \end{equation}
Consequently, this integral may be solvable and finite. Note that, an attempt of integrating the vacuum energy density  in the cosmic string spacetime will provide an infinite result since it is an idealized cosmic string, without structure \cite{BezerradeMello2011nv, MotaDispiration}. 

Let us now make use of Eq. \eqref{GZeta} to obtain an expression for the vacuum free energy. In order to do that we should apply the operator in Eq. \eqref{Box} on the scalar quantum field written as
  \begin{equation}\label{sol}
       \phi_j(x)=e^{-i\omega_n\tau}\varphi_{\ell}({\bf r}),
    \end{equation}
where $\varphi_{\ell}({\bf r})$ is the spatial parte of the solution of the scalar quantum field $ \phi_j(x)$, and $j=(n,\ell)$ are the quantum modes. One should notice that the present approach for thermal corrections is more convenient when we have an ultrastatic spacetime \cite{sonego2010, PontualMoraes, Cognola:1993qg, Bytsenko:1992hh}, in which case we can write the solution in the form of Eq. \eqref{sol}.

Thereby, the periodicity condition in Eq. \eqref{period} provides
   \begin{equation}\label{DR}
        \omega_n^2 = \left(\frac{2\pi n}{\beta}\right)^2\qquad\Longrightarrow\qquad \lambda_n = \omega_n^2 + \bf{k}^2,
    \end{equation}
with {\bf k} being the continuum momenta associated with the spatial coordinates and $n=0,\pm1,\pm 2,...$\;.

Upon substituting \eqref{sol}-\eqref{DR} in the expression \eqref{GZeta} we are able to note that the term $n=0$ can be extracted from the sum and, consequently, solved with the help of the identity \cite{Elizalde1994book}
\begin{equation}
        \sum_{n=-\infty}^\infty e^{-\eta n^2} = \sqrt{\frac{\pi}{\eta}}\sum_{n=-\infty}^\infty e^{-\frac{\pi^2 n^2}{\eta}}.
    \end{equation}
The zeta function \eqref{GZeta} is, then, given by
    \begin{equation} \label{zetaMod}
        \zeta_4(s) = \frac{\beta}{\sqrt{4\pi}\Gamma(s)}\left\{\Gamma(s - 1/2)\zeta_3(s-1/2) + 2\sum_{n=1}^\infty \int^\infty_0 \eta^{s-\frac{3}{2}}e^{-\frac{(n\beta)^2}{4\eta}}\text{Tr}\left[e^{-\eta\hat{A}_3}\right]d\eta\right\},
    \end{equation}
where the operator $\hat{A}_3$ is given by Eq. \eqref{Box}, without the time derivative. That is, it is only associated with the spatial part of the derivatives. Note that the zeta function $\zeta_3(s)$ is given by Eq. \eqref{GZeta} replacing $\hat{A}_4$ with $\hat{A}_3$. Furthermore, by using Eq. \eqref{Zlog}, with \eqref{zetaMod}, the free energy is found to be 
  \begin{eqnarray} \label{freeEnergy}
        F = - \frac{\ln Z}{\beta}= \frac{1}{2}\zeta_3(-1/2) - \frac{1}{\sqrt{4\pi}}\sum_{n=1}^\infty \int^\infty_0 \eta^{-\frac{3}{2}}e^{-\frac{(n\beta)^2}{4\eta}}\text{Tr}\left[e^{-\eta\hat{A}_3}\right]d\eta.
    \end{eqnarray}

We should now remember that the total quantities are obtained from the local quantities by integrating them in the space coordinates. Keeping that in mind, the free energy density \eqref{freeEnergy} is now written as
  \begin{equation}\label{freeEnergyDensity}
        \mathcal{F} = \frac{1}{2}\zeta_3({\bf r};-1/2) - \frac{1}{\sqrt{4\pi}}\sum_{n=1}^\infty\int_0^\infty \eta^{-\frac{3}{2}}e^{-\frac{(n\beta)^2}{4\eta}}K({\bf r},{\bf r},\eta)d\eta,
    \end{equation}
where we have used Eqs. \eqref{localZeta} and \eqref{traceK}. Additionally, the heat kernel $K({\bf r},{\bf r},\eta)$ can be calculated by using Eq. \eqref{HK}, with $\varphi_{\ell}({\bf r})$, and replacing $\lambda_j$ with ${\bf k^2}$.

The free energy density in Eq. \eqref{freeEnergyDensity} can be written more simply as 
  \begin{equation}\label{freeEnergyDensity2}
        \mathcal{F} = \mathcal{E}^0 + \Delta\mathcal{F},
    \end{equation}
 where $\mathcal{E}^0$ is the Casimir energy density at zero temperature given by the first term on the r.h.s of Eq. \eqref{freeEnergyDensity}, and $ \Delta\mathcal{F}$ is its temperature corrections given by the second term.
 
 Once we obtain the free energy density \eqref{freeEnergyDensity} we may be able to calculate thermodynamics quantities, namely, internal energy, entropy and pressure \cite{Landau}. In order to calculate the internal energy one should consider
   \begin{equation}\label{internal}
        U = -T^2\frac{\partial}{\partial T}\left(\frac{F}{T}\right).
    \end{equation}
 As to the entropy and pressure we have, respectively, 
   \begin{equation}\label{entropy}
        S = -\frac{\partial F}{\partial T},
    \end{equation}
  and 
      \begin{equation}\label{pressure}
        P = -\frac{\partial F}{\partial V},
    \end{equation} 
  which is calculated by taking the derivative of the free energy with respect to the volume $V$.
     
The generalized zeta function method, reviewed  in this section, to calculate the vacuum free energy can be found in more details in Refs.\cite{Hawking1977, Elizalde2012zv, Elizalde1994book}. We shall use it in the next sections to calculate temperature corrections to the scalar vacuum energy at zero temperature in Minkowski (Euclidean) spacetime under a helix boundary condition \cite{QuantumSpringFromCasimir, QuantumSpring, QuantumSpringD+1}.
%
\section{Scalar field under a helix boundary condition}
In this section we consider the scalar field quantum modes propagating in the (3+1)-Euclidean spacetime under the following helix boundary condition:
 \begin{equation}\label{HBC}
    \varphi(x+a,y,z) = \varphi(x,y+h,z),
 \end{equation}
where we have only considered the spatial part of the scalar field given by Eq. \eqref{sol}, $h$ is the pitch of the helix and $a$ its radius. Note that the helix boundary condition \eqref{HBC} for the scalar field have been previously considered in Refs. \cite{QuantumSpringFromCasimir, QuantumSpring, QuantumSpringD+1} in the context of the Casimir effect. Here we seek to consider temperature corrections to the Casimir energy density by using the generalized zeta function method behind the structure of Eq. \eqref{freeEnergyDensity} for the free energy density. The latter also allows us to obtain thermodynamics quantities such as internal energy, entropy and pressure.

The line element describing the (3+1)-Euclidean spacetime, in accordance with the formalism presented in the previous section, is given by
 \begin{equation}\label{LEE}
ds^2= -d\tau^2 - dx^2 - dy^2 - dz^2,
\end{equation}
where $\tau=it$ is the imaginary time and the range of the spacetime coordinates is $-\infty <t, x, y, z<\infty$. The line element in the form of Eq. \eqref{LEE} is necessary since the method for temperature corrections to the Casimir effect shown in the previous section is developed in the Euclidean spacetime, with imaginary time $\tau$.

\subsection{Equation of motion and heat kernel}
%
The equation of motion for the spatial part of the scalar field is also obtained by making use of Eq. \eqref{Box}, which in this case is for the operator $\hat{A}_3$ defined only by the spatial part of $\hat{A}_4$. The corresponding eigenvalues, $\lambda_{\ell}$, of the operator $\hat{A}_3$, once it acts on the scalar field, $\varphi_{\ell}({\bf r})$, provides an eigenvalue equation similar to the one in Eq. \eqref{EVE}, i.e.,
 \begin{equation}\label{Deq}
\left[- \frac{\partial^2}{\partial x^2} - \frac{\partial^2}{\partial y^2}- \frac{\partial^2}{\partial {z}^2} + m^2\right]\varphi_{\ell}({\bf r}) = \lambda_{\ell}\varphi_{\ell}({\bf r}),
\end{equation}
where $\ell$ stands for the spatial quantum modes. Note that, the helix structure of the boundary condition is defined such that $0\leq x \leq a $, $-h \leq  y < 0$, $-L/2 \leq z \leq L/2$ \cite{QuantumSpringFromCasimir, QuantumSpring, QuantumSpringD+1}.

 The general solution for the equation of motion \eqref{Deq}, under the helix boundary condition \eqref{HBC} is given by \cite{QuantumSpringFromCasimir, QuantumSpring, QuantumSpringD+1}
 \begin{eqnarray}\label{sol2}
 \varphi(x,y,z) &=& A e^{ik_x x+ik_y y + ik_z z}\nonumber\\
 &=&A e^{ik_x\left(x+\frac{ay}{h}\right)}  e^{ik_zz}e^{-2\pi in\frac{y}{h}},            
 \end{eqnarray}
where $A$ is a normalization constant to be determined and $ k_xa - k_yh = 2\pi n$, with $n=0,\pm1,\pm2, \pm3,...$\;.The corresponding eigenvalues are, thus, found to be
\begin{eqnarray}\label{rel}
\lambda_{\ell} &=& k_x^2 + \left(\frac{k_xa}{h} - \frac{2\pi n}{h}\right)^2 + k_{z}^2+ m^2\nonumber\\
&=& \left(\frac{k_yh}{a} + \frac{2\pi n}{a}\right)^2 + k_y^2 + k_{z}^2+ m^2. 
\end{eqnarray}
Since $k_x$ and $k_y$ are not independent momenta, we can see now that the spatial quantum modes are defined either as $\ell = (n,k_y,k_z)$ or $\ell = (n,k_x,k_z)$. From now on, we shall make use of the latter.

In order to calculate the heat kernel we need to determine the normalization constant $A$. For this, let us make use of the completeness relation
\begin{eqnarray}\label{Crel}
\sum_{n=-\infty}^{\infty}\int dk_xdk_y \varphi_{\ell}({\bf r} )\varphi^*_{\ell}({\bf r} ') = \frac{1}{\sqrt{-g^{(3)}}}\delta^3({\bf r} - {\bf r}' ),
\end{eqnarray}
where $g^{(3)}$ is the determinant of the spatial part of the metric present in the line element \eqref{LEE}. Upon using the solution \eqref{sol2} in \eqref{Crel} the normalization constant is obtained as
\begin{equation}\label{NC}
           A = \frac{1}{\sqrt{h(2\pi)^{2}}}.
\end{equation}
Therefore, the complete normalized solution of the scalar field under the helix boundary condition \eqref{HBC} is given by Eq. \eqref{sol2}, with the normalization constant in Eq. \eqref{NC}. The complete normalized solution makes possible now to calculate the heat kernel.

We can calculate the heat kernel from Eq. \eqref{HK}, along with Eqs. \eqref{sol2} and \eqref{NC}. This provides
        \begin{equation}\begin{split}\label{HKH1}
            K ({\bf r},{\bf r}',\eta) 
              & =  \frac{1}{(2\pi)^2h}e^{-m^2\eta} \sum_{n=-\infty}^\infty e^{-2\pi ni\frac{\Delta y}{h}}  \int_{-\infty}^\infty dk_x\int_{-\infty}^\infty dk_z e^{ - k_x^2\eta - \left(\frac{k_x a}{h}-\frac{2\pi n}{h}\right)^2 \eta - k_z^2\eta}\\
              & \times e^{ik_x\Delta u+ ik_z\Delta z}\;,     
        \end{split}\end{equation}
where $\Delta u = \left(\Delta x + \frac{a}{h}\Delta y\right) $. Thereby, by performing the integrals in the independent momenta $k_x$ and in $k_z$, we obtain
        \begin{equation}\label{HKH2}
            K ({\bf r},{\bf r}',\eta) =  \frac{a}{4\pi\eta d}  e^{-m^2\eta - \frac{\Delta z^2}{4\eta} - \frac{a^2\Delta u^2}{4\eta d^2}}  \vartheta(w, r),
       \end{equation}
with $d=(a^2+h^2)^{\frac{1}{2}}$ and $\vartheta(w, r)$ is the Jacobi theta function defined as
    \begin{equation}\label{TF}
            \vartheta(w, r) = \sum_{n=-\infty}^\infty e^{i\pi n^2 r - 2\pi i n w}.
       \end{equation}
Note that in our case, $w=\frac{(h\Delta y - a\Delta x)}{d^2}$  and $r = \frac{4\pi\eta i}{d^2}$. In order to obtain the Euclidean heat kernel contribution from \eqref{HKH2} we can make use of the identity \cite{Elizalde1994book}
 \begin{equation}\label{TFI}
            \vartheta(w, r) = \frac{1}{\sqrt{-i r}}e^{-i\pi\frac{w^2}{r}} \vartheta\left(\frac{w}{r}, -\frac{1}{r}\right).
       \end{equation}    
Consequently, the heat kernel obtained in Eq. \eqref{HKH2} becomes
      \begin{equation}\label{HKH3}
            K ({\bf r},{\bf r}',\eta) =  \frac{1}{(4\pi \eta)^{\frac{3}{2}}}     e^{-m^2\eta - \frac{\Delta z^2}{4\eta} - \frac{a^2\Delta u^2}{4\eta d^2}-\frac{a^2\left(\Delta y - \frac{a}{h}\Delta x\right)^2}{4\eta d^2}}\sum_{n=-\infty}^\infty e^{-\frac{n^2 d^2}{4\eta} - \frac{n (h\Delta y - a\Delta x)}{2\eta}}.
       \end{equation}
We can now proceed to take the coincidence limit, ${\bf r}'\rightarrow {\bf r}$, in the above two-point function heat kernel of the scalar field, $\varphi_{\ell}({\bf r})$, under the helix boundary condition \eqref{HBC}. This gives
       \begin{equation}\label{HKH41}
            K ({\bf r},{\bf r},\eta) =  \frac{e^{-m^2\eta}}{(4\pi \eta)^{\frac{3}{2}}}\sum_{n=-\infty}^\infty e^{-\frac{n^2 d^2}{4\eta}}.
       \end{equation}      
It is straightforward to see that the Euclidean heat kernel is obtained from the term $n=0$ \cite{Cognola:1993qg, Bytsenko:1992hh, PontualMoraes}, that is,
   \begin{equation}\label{HKE}
            K_{\text{E}} ({\bf r},{\bf r},\eta) =  \frac{1}{(4\pi \eta)^{\frac{3}{2}}}   e^{-m^2\eta}.
       \end{equation}
The Euclidean heat kernel above gives a divergent contribution for the integral in $\eta$ present in the zeta function definition in Eq. \eqref{localZeta} and, therefore, should be dropped in order to obtain a renormalized Casimir energy density. Moreover, the contribution of the Euclidean heat kernel \eqref{HKE} for the second term on the r.h.s of the free energy density \eqref{freeEnergyDensity} gives the scalar thermal (blackbody) radiation contribution \cite{Hawking1977}. As it is known, the latter should be subtracted providing a finite renormalization for the thermal correction contributions \cite{PhysRevD.83.104042, Bezerra:2011nc}. This is necessary if we want to obtain the correct classical limit at high temperatures for the free energy density, as we shall see later.

The subtraction of the Euclidean heat kernel contribution \eqref{HKE} from \eqref{HKH41} allows us to obtain the renormalized heat kernel for the scalar field, $\varphi_{\ell}({\bf r})$, under the helix boundary condition \eqref{HBC}, i.e.,
\begin{equation}\label{HKH4}
            K_{\text{ren}} ({\bf r},{\bf r},\eta) =  \frac{e^{-m^2\eta}}{4(\pi \eta)^{\frac{3}{2}}}     \sum_{n=1}^\infty e^{-\frac{n^2 (a^2 + h^2)}{4\eta}}.
\end{equation}         

 Let us now turn to the calculation of the free energy density defined in Eq. \eqref{freeEnergyDensity}. The first term on the r.h.s gives the contribution to the Casimir energy density at zero temperature, and can be obtained by substituting the heat kernel \eqref{HKH4} in \eqref{localZeta} for $s=-\frac{1}{2}$. This gives
 \begin{equation}\label{LZE}
   \mathcal{E}_{\text{ren}} =\frac{1}{2}\zeta_3({\bf r},-1/2) = -\frac{m^4}{2\pi^2}\sum_{n=1}^\infty f_2(mnd),
\end{equation}        
where 
 \begin{equation}\label{func1}
  f_{\mu}(x) = \frac{K_{\mu}(x)}{x^{\mu}},
\end{equation} 
with $K_{\mu}(x)$ being the modified Bessel function of the second kind, also known as the Macdonald function. The expression in Eq. \eqref{LZE} for the renormalized Casimir energy density at zero temperature has been obtained previously by the authors in Refs \cite{QuantumSpringFromCasimir, QuantumSpring, QuantumSpringD+1}. So, our result is consistent with the one from the latter.

The massless scalar field renormalized Casimir energy density can be obtained from \eqref{LZE} in the limit of small arguments for the Macdonald function, i.e., $K_{\mu}(x)\simeq \frac{\Gamma(\mu)}{2}\left(\frac{2}{x}\right)^{\mu}$ \cite{abramowitz, gradshteyn2000table}. Thus, we have \cite{QuantumSpringFromCasimir, QuantumSpring, QuantumSpringD+1}
 \begin{equation}\label{LZEMassless1}
   \mathcal{E}^0_{\text{ren}} = -\frac{\pi^2}{90d^4},
\end{equation}  
where we have used the Riemann zeta function $\zeta(4)=\frac{\pi^2}{90}$ \cite{Elizalde1994book, abramowitz, gradshteyn2000table}. A brief discussion on whether the energy density \eqref{LZEMassless1} would depend on the choice of regularization method can be found on Appendix A.

We want now to calculate temperature corrections to the renormalized Casimir energy densities in Eqs. \eqref{LZE} and \eqref{LZEMassless1} for the massive and massless scalar field cases, respectively. Before doing that, let us remind that although the Euclidean heat kernel provides a divergent contribution to the Casimir energy densities \eqref{LZE} and \eqref{LZEMassless1}, it gives a finite contribution to the temperature correction expression on the second term on the r.h.s of Eq. \eqref{freeEnergyDensity2}. By Substituting the Euclidean heat kernel \eqref{HKE} in the expression for $\Delta\mathcal{F}$ present in Eq. \eqref{freeEnergyDensity2} we obtain 
 \begin{equation}\label{EUT}
  \Delta\mathcal{F}_{\text{E}}= -\frac{m^4}{2\pi^2}\sum_{n=1}^\infty f_2(mn\beta),\qquad\qquad\qquad   \Delta\mathcal{F}_{\text{E}}= -\frac{\pi^2}{90}(k_BT)^4,
\end{equation}   
for the massive and massless scalar fields, respectively. The expression on the r.h.s of Eq. \eqref{EUT} is the massless scalar thermal (blackbody) radiation contribution \cite{Hawking1977}. This term should be subtracted from $\Delta\mathcal{F}$ in order to obtain, at high temperature, the correct classical limit \cite{PhysRevD.83.104042, Bezerra:2011nc}. This is in fact a finite renormalization since the expressions  in \eqref{EUT} to be subtracted are not divergent.

The renormalized temperature correction, $ \Delta\mathcal{F}_{\text{ren}}$, to the Casimir energy density \eqref{LZE} is calculated by using the renormalized heat kernel \eqref{HKH4}. This provides
 \begin{equation}\label{LZETC}
  \Delta\mathcal{F}_{\text{ren}} = -\frac{m^4}{\pi^2}\sum_{j=1}^{\infty}\sum_{n=1}^\infty f_2\left[m\beta(j^2 + n^2\gamma^2)^{\frac{1}{2}}\right],
\end{equation}   
where $\gamma = \frac{d}{\beta}=k_BT d$. This expression is exponentially suppressed for large arguments, $\beta m\gg 1$, of the Macdonald function, i.e., $K_{\mu}(x)\simeq \sqrt{\frac{\pi}{2x}}e^{-x}$ \cite{abramowitz, gradshteyn2000table}. This is in fact the limit of low temperatures, $\frac{k_BT}{m}\ll 1$, for the expression in Eq. \eqref{LZETC}, in accordance with the fact that the temperature corrections must vanish in this regime. On the other hand, for small arguments, $\beta m\ll 1$, we obtain the massless expression from the massive thermal correction \eqref{LZETC}, that is,
 \begin{equation}\label{Massless}
  \Delta\mathcal{F}_{\text{ren}} = -\frac{2}{\pi^2}\sum_{n=1}^\infty\sum_{j=1}^\infty \frac{1}{(j^2\beta^2+n^2d^2)^{2}}.
\end{equation}   

From Eq. \eqref{freeEnergyDensity2}, the renormalized free energy density is written as the sum of Eqs. \eqref{LZE} and \eqref{LZETC} for the massive case. This provides
 \begin{equation}\label{Massive}
 \mathcal{F}_{\text{ren}} = -\frac{m^4}{2\pi^2}\sum_{n=1}^\infty f_2(mnd) - \frac{m^4}{\pi^2}\sum_{j=1}^{\infty}\sum_{n=1}^\infty f_2\left[m\beta(j^2 + n^2\gamma^2)^{\frac{1}{2}}\right].
\end{equation} 
Again, in the limit of large arguments $m\beta\gg 1$, the second term on the r.h.s is exponentially suppressed so that only the first term, at zero temperature, survives. Thus we recover the result obtained in Refs.\cite{QuantumSpringFromCasimir, QuantumSpring, QuantumSpringD+1}.

As to the massless case, the sum of Eqs. \eqref{LZEMassless1} and \eqref{Massless} is written as 
 \begin{equation}\label{MasslessFE}
 \mathcal{F}_{\text{ren}} = -\frac{\pi^2}{90d^4} - \frac{2}{\pi^2}\sum_{n=1}^\infty\sum_{j=1}^\infty \frac{1}{(j^2\beta^2+n^2d^2)^{2}}.
\end{equation}  
The temperature corrections present in the free energy density expressions \eqref{Massive} and \eqref{MasslessFE}, to the best of our knowledge, are new results obtained for the first time here.

In particular, for the massless case, the free energy density \eqref{MasslessFE} can be further developed in order to obtain the limits of high and low temperatures. For this purpose, to perform the sum in $j$ first  in Eq. \eqref{Massive} allows us to obtain the high-temperature limit of the expression \eqref{MasslessFE} whilst to perform the sum in $n$ first allows us to obtain its low-temperature limit. We shall do this analysis next.
%
\subsection{High-temperature limit}
%
We wish now to consider the high-temperature limit of the free energy density \eqref{MasslessFE} for the massless scalar field. In order to do that, it is convenient to perform first the sum in $j$ present in \eqref{Massless}. This can be done by writing the latter in terms of the Epstein-Hurwitz zeta function $\zeta_{EH}(s,M)$, providing
 \begin{equation}\label{LZETCMassless}
  \Delta\mathcal{F}_{\text{ren}} = -\frac{2}{\pi^2\beta^4}\sum_{n=1}^\infty\zeta_{EH}(s,n\gamma),
\end{equation}   
where $\gamma = \frac{d}{\beta}$ and
        \begin{equation}\label{zetaEpsteinHurwitzDef}
            \zeta_{EH}(s,M) = \sum_{k=1}^\infty (k^2+M^2)^{-s}.
        \end{equation}
 The Epstein-Hurwitz zeta function above is defined for Re($s$) $> \frac{1}{2}$ and $M^2\ge 0$ \cite{Elizalde1994book}. Note that we recover \eqref{MasslessFE} by taking the limit $s\rightarrow2$ in Eq. \eqref{LZETCMassless}. However, before doing that, let us make use of the following analytic continuation for \eqref{zetaEpsteinHurwitzDef} \cite{Elizalde1994, Elizalde1994book, Bezerra:2017zqq}:
        \begin{equation}\label{zetaEpsteinHurwitzExt}
            \zeta_{EH}\left(s, M\right) = -\frac{M^{-2s}}{2} + \frac{\sqrt{\pi}}{2}\frac{\Gamma(s-1/2)}{\Gamma(s)}M^{(1-2s)} 
            + \frac{2^{\frac{3}{2}-s}\pi^{\frac{1}{2}} M^{1-2s}}{\Gamma(s)}\sum_{k=1}^{\infty}f_{(1/2-s)}(2\pi kM),
        \end{equation}
which is valid for other values of $s$. One should remind that the function $f_{\mu}(x)$ has been defined, in Eq. \eqref{func1}, in terms of the Macdonald function $K_{\mu}(x)$, and $\Gamma(x)$ is the gamma function. Thereby, by substituting \eqref{zetaEpsteinHurwitzExt} in \eqref{LZETCMassless}, taking the limit $s\rightarrow2$ and performing the sum in $j$, we obtain 
 \begin{equation}\label{LZETCMasslessj}
  \Delta\mathcal{F}_{\text{ren}} = \frac{\pi^2}{90d^4} - \frac{k_BT}{2\pi d^3}\zeta(3) - \frac{k_BT}{4\pi d^3}\sum_{n=1}^\infty n^{-3}\left[\frac{2 e^{-\pi n\gamma}\sinh(\pi n\gamma) + 2\pi n\gamma}{\sinh^2(\pi n\gamma)}\right], 
   \end{equation}   
where $\zeta(s)$ is the Riemann zeta function \cite{Elizalde1994book}. Furthermore, By using \eqref{LZETCMasslessj}, the free energy density \eqref{MasslessFE} may now be written as 
 \begin{equation}\label{FER}
 \mathcal{F}_{\text{ren}} = - \frac{k_BT}{2\pi d^3}\zeta(3) - \frac{k_BT}{4\pi d^3}\sum_{n=1}^\infty n^{-3}\left[\frac{2 e^{-\pi n\gamma}\sinh(\pi n\gamma) + 2\pi n\gamma}{\sinh^2(\pi n\gamma)}\right]. 
  \end{equation}   
 The expression above for the free energy density is more convenient to provide the high-temperature limit $\gamma\gg 1$ or, putting in another way, $k_BTd\gg 1$. In this regime we have
 \begin{equation}\label{FERHigh}
 \mathcal{F}_{\text{ren}} \simeq - \frac{k_BT}{2\pi d^3}\zeta(3) - \frac{2(k_BT)^2}{d^2}e^{-2\pi\gamma}, 
 \end{equation}   
 where the second term on the r.h.s is exponentially suppressed while the first one is the classical limit, which dominates, as it should be at high temperatures. We should note that, had we considered the scalar thermal radiation contribution in Eq. \eqref{EUT} the free energy density \eqref{FERHigh}, at high temperatures, would not have given correctly only the classical limit \cite{PhysRevD.83.104042, Bezerra:2011nc}. That is why the free energy density had to suffer a finite renormalization, by subtracting from it the term proportional to $(k_BT)^4$ coming from the scalar thermal radiation contribution.
 
 Additional thermodynamics quantities of interest can be calculated, such as internal energy density and entropy density. Thus, we firstly consider the calculation of the internal energy density by making use of Eq. \eqref{internal} and the renormalized free energy density \eqref{FER}. With this, we obtain 
 \begin{equation}\label{RIE} 
 \mathcal{U}_{\text{ren}} =  - \frac{\pi (k_BT)^3}{d} \sum_{n=1}^\infty \frac{\cosh(\pi n\gamma)}{n\sinh^3(\pi n\gamma)}.
  \end{equation}   
 Its corresponding high-temperature limit is exponentially suppressed, that is,
 \begin{equation}\label{RIEHigh}
 \mathcal{U}_{\text{ren}} \simeq  - \frac{4\pi (k_BT)^3}{d}e^{-2\pi\gamma}.
   \end{equation} 

As to the renormalized entropy density, it can be obtained by substituting the renormalized free energy density \eqref{FER} in \eqref{entropy}. This gives
 \begin{equation}\label{entropyR} 
 \mathcal{S}_{\text{ren}} = \frac{k_B}{2\pi d^3}\zeta(3) - \frac{k_B}{2\pi d^3} \sum_{n=1}^\infty n^{-3}e^{-n\pi\gamma}\left\{\frac{-1 - n \pi\gamma + 
 n \pi\gamma \coth(n \pi\gamma)\left[-1 + 2 n \pi\gamma + 2 n \pi\gamma \coth(n\pi\gamma)\right]}{\sinh(n\pi\gamma)}\right\}, 
  \end{equation}   
which at the high-temperature limit provides the dominant terms
 \begin{equation}\label{entropyRHigh}
 \mathcal{S}_{\text{ren}} \simeq \frac{k_B}{2\pi d^3}\zeta(3) - \frac{2\pi k_B}{d}(k_BT)^2e^{-2\pi\gamma}.
  \end{equation}  
  The high-temperature regime exhibited in the above expression for the entropy density is dominated by the first term on the r.h.s since the second term is exponentially suppressed.
  %
\subsection{Low-temperature limit}
%
Let us now turn to the low-temperature limit, that is, $k_BTd\ll 1$ ($\gamma\ll 1$). In order to obtain expressions in this regime, analogously with what we have done previously for the high-temperature limit,  we need now to solve first the sum in $n$ present in Eq. \eqref{Massless}. Thereby, by writing the latter in terms of the Epstein-Hurwitz zeta function $\zeta_{EH}(s,M)$, we have
 \begin{equation}\label{LZETCMasslesssu_n}
  \Delta\mathcal{F}_{\text{ren}} = -\frac{2}{\pi^2d^4}\sum_{j=1}^\infty\zeta_{EH}(s,j/\gamma).
\end{equation}  
In the expression above, we can now make use of the analytic extension \eqref{zetaEpsteinHurwitzExt} for Epstein-Hurwitz zeta function. After doing that, we take the limit $s\rightarrow 2$ and perform the sum in $n$. This gives
  \begin{equation}\label{LZETCMasslessn}
  \Delta\mathcal{F}_{\text{ren}} = \frac{\pi^2}{90}(k_BT)^4 - \frac{(k_BT)^3}{2\pi d}\zeta(3) - \frac{(k_BT)^2}{4\pi d^2}\sum_{j=1}^\infty j^{-3}\left[\frac{2\gamma e^{-\frac{\pi j}{\gamma}}\sinh(\pi j/\gamma) + 2\pi j}{\sinh^2(\pi j/\gamma)}\right],
  \end{equation} 
  for the renormalized temperature correction to the Casimir energy density \eqref{LZEMassless1} at zero temperature. Thus, the renormalized free energy density \eqref{MasslessFE}  becomes
   \begin{equation}\label{FERtotalsu_n1}
 \mathcal{F}_{\text{ren}} = -\frac{\pi^2}{90d^4}
+ \frac{\pi^2}{90}(k_BT)^4 - \frac{(k_BT)^3}{2\pi d}\zeta(3) - \frac{(k_BT)^2}{4\pi d^2}\sum_{j=1}^\infty j^{-3}\left[\frac{2\gamma e^{-\frac{\pi j}{\gamma}}\sinh(\pi j/\gamma) + 2\pi j}{\sinh^2(\pi j/\gamma)}\right].
  \end{equation}  
Furthermore, the low-temperature limit of the free energy density above is shown to have the form
     \begin{equation}\label{FERtotalsu_nlow}
 \mathcal{F}_{\text{ren}} \simeq -\frac{\pi^2}{90d^4}
+ \frac{\pi^2}{90}(k_BT)^4 - \frac{(k_BT)^3}{2\pi d}\zeta(3) - \frac{2(k_BT)^3}{\pi d}e^{-\frac{2\pi}{k_BTd}},
  \end{equation}  
 which is clearly dominated by the first term on the r.h.s, corresponding to the Casimir energy density at zero temperature. Note also that the last term on the r.h.s is exponentially suppressed. 
 
 Let us turn now to the calculation of both the renormalized internal energy density and entropy density. The former is obtained by using Eqs. \eqref{internal} and \eqref{FERtotalsu_n1}, providing 
 \begin{eqnarray}\label{RIEsu_n1}
 \mathcal{U}_{\text{ren}} &=&  -\frac{\pi^2}{90d^4} - \frac{\pi^2}{30}(k_BT)^4 +  \frac{(k_BT)^3}{\pi d}\zeta(3)\nonumber\\
& + &\frac{k_BT}{\pi d^3}\sum_{j=1}^\infty j^{-3}e^{-\frac{j\pi}{\gamma}}\left\{\frac{(j\pi + \gamma)\gamma + j\pi\coth(j\pi/\gamma)\left[j\pi + \gamma + j\pi\coth(j\pi/\gamma)\right]}{\sinh(j\pi/\gamma)}\right\}.
  \end{eqnarray} 
Its corresponding  low-temperature limit is found to be
  \begin{equation}\label{RIEsu_nlow}
 \mathcal{U}_{\text{ren}} \simeq  -\frac{\pi^2}{90d^4} - \frac{\pi^2}{30}(k_BT)^4 +  \frac{(k_BT)^3}{\pi d}\zeta(3) + \frac{2(k_BT)^3}{\pi d}e^{-\frac{2\pi}{k_BTd}}.
   \end{equation} 
 Once again, the Casimir energy density at zero temperature is the dominant term on the r.h.s, as it should be, whilst the last term is exponentially suppressed.
 
 Finally, we can now calculate the renormalized entropy density by using Eqs. \eqref{entropy} and \eqref{FERtotalsu_n1}. This allows us to obtain
   \begin{eqnarray}\label{entropyn}
 \mathcal{S}_{\text{ren}} &=& 
- \frac{2\pi^2k_B}{45}(k_BT)^3 + \frac{3k_B}{2\pi d}(k_BT)^2\zeta(3) \nonumber\\
&+& \frac{k_B}{2\pi d^3}\sum_{j=1}^\infty j^{-3}e^{-\frac{j\pi}{\gamma}}\left\{\frac{3\gamma(j\pi + \gamma) + j\pi\coth(j\pi/\gamma)\left[2j\pi + 3\gamma + 2j\pi\coth(j\pi/\gamma)\right]}{\sinh(j\pi/\gamma)}\right\}.
  \end{eqnarray}  
 Its low-temperature limit also contains an exponentially suppressed term, similarly to the others quantities, i.e.,
   \begin{eqnarray}\label{entropynlow}
 \mathcal{S}_{\text{ren}} \simeq
- \frac{2\pi^2k_B}{45}(k_BT)^3 + \frac{3k_B}{2\pi d}(k_BT)^2\zeta(3) + \frac{3k_B}{\pi d}(k_BT)^2e^{-\frac{2\pi}{k_BTd}}.
  \end{eqnarray}   
  One should note that the entropy density associated with the massless scalar field under a helix boundary condition vanishes when the temperature goes to zero, in accordance with the third law of thermodynamics (the Nernst heat theorem)\cite{LandauStatistical, Bezerra:2011nc, PhysRevD.83.104042}. One should also note that although we have considered the convenient expressions \eqref{FER}, \eqref{RIE} and \eqref{entropyR} to obtain the high-temperature limit and the expressions \eqref{FERtotalsu_n1}, \eqref{RIEsu_n1} and \eqref{entropyn} to obtain the low-temperature limit, they are all equivalent but only expressed in different forms.
 %
 \subsection{Pressure and equation of state}
%
Let us now consider the massless scalar field expression for the renormalized free energy density in Eq. \eqref{MasslessFE} to calculate the pressure and show that it satisfies an equation of state. We, then, start by defining the renormalized free energy from Eq. \eqref{MasslessFE}  as
 \begin{eqnarray}\label{Masslessrfreeenergy}
 F_{\text{ren}}&=&d^3 \mathcal{F}_{\text{ren}} \nonumber\\
 &=& E^0_{\text{ren}} -\frac{2}{\pi^2d}f(\gamma),
\end{eqnarray}   
where $V=d^3$ is considered as being the volume and $E^0_{\text{ren}}$ is the renormalized Casimir energy given by 
   \begin{equation}\label{LZEMassless}
  E^0_{\text{ren}} = -\frac{\pi^2}{90d}.
\end{equation} 
Additionally, we define the function $f(\gamma)$ as
 \begin{equation}\label{function}
f(\gamma) = \sum_{n=1}^\infty\sum_{j=1}^\infty \left(\frac{j^2}{\gamma^2}+n^2\right)^{-2}.
\end{equation} 
Consequently, the pressure can be obtained from Eq. \eqref{pressure}, which we can write in the form
 \begin{eqnarray}\label{pressure1}
P_{\text{ren}} &=&- \frac{\partial F_\text{ren}}{\partial V}\nonumber\\
&=&-\frac{1}{3d^2}\frac{\partial F_\text{ren}}{\partial d}.
\end{eqnarray} 
By substituting the renormalized free energy \eqref{Masslessrfreeenergy} in the above equation we have 
 \begin{eqnarray}\label{pressure2}
P_{\text{ren}} = \frac{1}{3}\mathcal{E}^0_{\text{ren}} -\frac{2}{3\pi^2d^4}f(\gamma) + \frac{2k_BT}{3\pi^2d^3}\frac{\partial f(\gamma)}{\partial\gamma}.
\end{eqnarray} 
Note that this expression is a closed form for the pressure. We could still go further and perform either the sum in $j$ or in $n$ present in the function $f(\gamma)$, as we have previously done, and develop even more the expression in Eq. \eqref{pressure2}. However, we want to focus here in showing that the renormalized pressure obeys an equation of state. 

The internal energy density can alternatively be obtained by using Eqs. \eqref{internal} and \eqref{MasslessFE}, that is,
 \begin{eqnarray}\label{RIEsu_n11}
 \mathcal{U}_{\text{ren}} &=& -T^2\frac{\partial}{\partial T}\left[\frac{\mathcal{F}_\text{ren}}{T}\right]\nonumber\\
& =& \mathcal{E}^0_{\text{ren}} - \frac{2}{\pi^2d^4}f(\gamma) + \frac{2k_BT}{\pi^2d^3}\frac{\partial f(\gamma)}{\partial\gamma}.
  \end{eqnarray} 
By comparing Eqs. \eqref{pressure2} and \eqref{RIEsu_n11} we can show that
 \begin{eqnarray}\label{RIEsu_n111}
P_{\text{ren}} = \frac{1}{3} \mathcal{U}_{\text{ren}}, 
\end{eqnarray}
which is the wanted equation of state.

It is clear that the generalized zeta function method to calculate the Casimir energy density, as well as its temperature corrections, is an elegant and precise method. The latter has allowed us to obtain temperature corrections to the Casimir energy density associated with the scalar field under a helix boundary condition, along with the high- and low-temperature limits. Although the Casimir energy densities \eqref{LZE} and \eqref{LZEMassless1} have previously been obtained in Refs. \cite{QuantumSpringFromCasimir, QuantumSpring, QuantumSpringD+1}, their temperature corrections along with the high- and low-temperature limits for the massless scalar field case have been obtained, to the best of our knowledge, for the first time here. We would like to stress that the importance of conducting the present investigation of the Casimir effect, with temperature corrections, relies on the fact that a helix geometry can represent an DNA structure and cell membrane proteins. Furthermore, the need for improvements in nano-electromechanical and micro-electromechanical systems, along with applications in nanotubes, reinforces the importance of studying, as a whole, quantum vacuum fluctuation effects of Casimir-type.
%
\section{Conclusions}
%
In this work we have overviewed how the partition function can be obtained from the path integral for the scalar field and how it is connected with the generalized zeta function by means of Eq. \eqref{Zlog}. By linking the generalized zeta function with the partition function we have been able to obtain the closed expressions \eqref{freeEnergy} and \eqref{freeEnergyDensity} for the free energy and free energy density, respectively. These expressions explicitly show that they are composed by a term for the zero temperature obtained from the usual calculations for the vacuum energy, and a term for temperature corrections - these are important expressions since experiments are normally performed at a finite temperature. The equation for the free energy density has allowed us to obtain regularized thermodynamical quantities from the partition function such as the free energy, pressure, entropy and internal energy.

To further our analysis, by making use of the generalized zeta function method we have considered the nontrivial topology of the helix boundary condition in a (3+1)-dimensional Euclidean spacetime. We were able to obtain the renormalized heat kernel \eqref{HKH4} and, upon taking the coincidence limit we could see that it produces a sum with the $n=0$ term corresponding to the Euclidean contribution \eqref{HKE}. When applied to the free energy density at zero temperature the Euclidean heat kernel produces a divergent term and for that, it should be subtracted. On the other hand, for the thermal corrections, the Euclidean heat kernel, gives a finite contribution to the massive scalar field and the blackbody radiation contribution for the massless scalar field (see Eq. \eqref{EUT}). Although these contributions are finite we have shown that they should also be subtracted in order to obtain the correct classical limit at high temperature for the massless scalar field case and, hence, occurring a finite renormalization of the free energy density. The renormalization procedure for both the vacuum energy at zero temperature and for its thermal corrections allowed us to obtain closed and analytical expressions for the renormalized free energy densities \eqref{Massive} and \eqref{MasslessFE} for the massive and massless scalar fields, respectively. 

We have also shown that the massless free energy density \eqref{MasslessFE} has a particularity: for asymptotic behaviours with respect to the temperature, the resulting free energy density is conveniently expressed to analyze the high- and low-temperature limits depending on which index of the double sum is performed first. Hence, by performing first the sum over $j$ we have been able to obtain the renormalized free energy density \eqref{FER} and its high-temperature limit \eqref{FERHigh}. With this asymptotic expression we have seen that it provides the correct classical limit, proportional to $k_BT$, which would not be possible had we considered the blackbody scalar radiation contribution \eqref{MasslessFE}. The thermodynamics quantities, namely, internal energy density \eqref{RIE}, entropy density \eqref{entropyR} and their corresponding high-temperature limits \eqref{RIEHigh} and \eqref{entropyRHigh} have also been calculated.

On the other hand, by performing first the sum over $n$ in Eq. \eqref{MasslessFE} we have been able to obtain the renormalized free energy density \eqref{FERtotalsu_n1} and its low-temperature limit \eqref{FERtotalsu_nlow}, providing that the dominant term is the vacuum energy at zero temperature. We have also obtained the internal energy density \eqref{RIEsu_n1}, entropy density \eqref{entropyn} and their corresponding low-temperature limits \eqref{RIEsu_nlow} and \eqref{entropynlow}. These asymptotic expressions showed that the internal energy density provide the vacuum energy density at zero temperature as the dominant term, and that the entropy goes to zero as the temperature vanishes, in accordance with the third law of thermodynamics. We have pointed out that Eqs. \eqref{FER}, \eqref{RIE}, \eqref{entropyR} and \eqref{FERtotalsu_n1}, \eqref{RIEsu_n1}, \eqref{entropyn} are all equivalent. They are only expressed in different and convenient forms, in order to analyze the high- and low-temperature limits.

Moreover, in order to show that the pressure obeys an equation of state, we have defined the free energy \eqref{Masslessrfreeenergy} and calculated the pressure in Eq. \eqref{pressure2}. With this, it was straightforward to compare the latter with the internal energy density expression \eqref{RIEsu_n11} to obtain the equation of state \eqref{RIEsu_n111}.

It is interesting to note that, as a starting point approximation, one could use the Casimir energy density \eqref{LZEMassless1}, at zero temperature, to see whether it can significantly induce genetic mutations on the DNA or RNA. The DNA is formed by a double sugar-phosphate strand linked to one another by hydrogen bonds of its nitrogen bases that shapes the DNA in something like a twisted ladder. For the most common type of DNA, known as the B-DNA, the pitch of each single strand structure is approximately $3.4nm$ containing ten nucleotides with $0.34nm$ separation and about $2nm$ distance between each strand \cite{sinden1994dna}. Those are exactly the length scales where the Casimir effect may play a significant role. As for the RNA, the single strand length can vary through a wide range of nucleotide number \cite{RNASize, RNASmallSize}. Hence, this data can be used to estimate the Casimir energy density in a cylinder-unit of DNA or RNA. In order to also estimate the chance of a mutation to occur as a consequence of the Casimir energy density \eqref{LZEMassless1} a cutoff energy should be adopted. In this sense, the x-ray radiation energy, which may induce genetic mutations, can be used for this purpose. Of course, DNA and RNA are not found in vacuum neither influenced only by the scalar field vacuum state, since there are other quantum fields that might influence the results. Also, the environmental variables such as the composition of the cell, organelles and surrounding structures should be taken into account to see whether the Casimir effect arising from the helix condition will be able to significantly induce genetic mutations or not. However, a simple application of the Casimir energy density \eqref{LZEMassless1}, as described initially,  can provide a good first approximation and open news ways to see how we could take into account other important components which might influence the occurrence of genetic mutations. This is in fact, a work in progress.

{\acknowledgments}
The authors thank Celio Muniz for helpful discussions. The author G.A. is funded through an undergraduate scholarship by the Universidade Federal da Para\'iba (UFPB) - PIBIC program. The author H.F.S.M. is supported by the Brazilian agency CNPq under grants 305379/2017-8 and 311031/2020-0.
\\
\\

\appendix 
\section{The Cutoff Method}
The zeta function method presented in this paper directly provides a way of regularize and, consequently, renormalize the vacuum energy density, as seen in the calculation of Eq. \eqref{LZEMassless1}, and apparently  does not take into account the energy outside the boundary, that is, the pure Minkowski spacetime contribution. This happens as a consequence of the analytical continuation property that the zeta function possesses \cite{Elizalde:2020mlv}. We want now, however, to check whether the calculation of the vacuum energy density \eqref{LZEMassless1} would depend on the regularization method, as it does, for instance, on the case of the rectangular box when one makes use of the cutoff method \cite{Cavalcanti:2003tw, Edery:2005bx}. We will show below, for the massless scalar field case, that this does not seem to be the case for the helix boundary condition.

The vacuum energy density, at zero temperature, for the massless scalar field, in $(3+1)$-dimensional spacetime, can be calculated by considering the eigenfrequencies \eqref{rel} as \cite{BordagMohideenMostepanenko, QuantumSpring,QuantumSpringD+1,QuantumSpringFromCasimir}

\begin{equation}\label{A1}
    \mathcal{E}_0 = \frac{1}{2h}\int \frac{dk_x\:dk_z}{(2\pi)^2}\sum_{n=-\infty}^{\infty}\sqrt{k_x^2 + \left(\frac{k_xa}{h} - \frac{2\pi n}{h}\right)^2 + k_{z}^2}.
\end{equation}
Next, as it is usually done when one uses the cutoff method, we must introduce in the integral above a damping function of the form $e^{-\delta\sqrt{k_x^2 + \left(\frac{k_xa}{h} - \frac{2\pi n}{h}\right)^2 + k_{z}^2}}$, where $\delta$ is a regularization parameter with dimension of length. This is necessary in order to make both the integral and summation to converge in Eq. \eqref{A1}. After introducing the damping function we are able first to solve the integral and last the sum, leading to the regularized vacuum energy density 
\begin{equation}\begin{split}\label{regE}
    \mathcal{E}_0(\delta) = & \frac{1}{\pi d^3 \delta^3 (e^{\frac{2\pi\delta}{d}}-1)^3}\left[\cosh\left(\frac{3\pi\delta}{d}\right)+\sinh\left(\frac{3\pi\delta}{d}\right)\right] \\
    & \quad \times \left[d^2\cosh\left(\frac{3\pi\delta}{d}\right)-(d^2-4\pi^2\delta^2)\cosh\left(\frac{\pi\delta}{d}\right)+4\pi d\delta\sinh\left(\frac{\pi\delta}{d}\right)\right], 
\end{split}\end{equation}
where $d = \sqrt{a^2+h^2}$. Note that the vacuum energy density should not depend, in principle, on the regularization parameter $\delta$ so that we must take the limit $\delta\rightarrow 0$ at some point. However, before taking the limit $\delta\rightarrow 0$ in the expression above we must first identify the divergent contribution by expanding \eqref{regE} in series for small values of $\delta$, that is,
\begin{equation}\label{appendixEnDen}
     \mathcal{E}_0(\delta) = \frac{3}{2\pi^2\delta^4}-\frac{\pi^2}{90d^4}+\frac{2\pi^4\delta^2}{315d^6}+\mathcal{O}(\delta^4).
\end{equation}
We should note that second order terms, or higher, in $\delta$ go to zero as we remove the damping factor ($\delta\rightarrow 0$). Moreover, the first term on the r.h.s of Eq. \eqref{appendixEnDen} is the divergent contribution and should be removed as we shall show below. It is in fact the Minkowski contribution, the one outside the boundary. The second term as we can notice, is the finite vacuum energy density obtained in Eq. \eqref{LZEMassless1}.

Let us show that the first term on the r.h.s of Eq. \eqref{appendixEnDen} is in fact the Minkowski contribution. Consider then the expression \eqref{A1} but with all continuum momenta \cite{BordagMohideenMostepanenko}, i.e.,
\begin{equation}
    \mathcal{E}_{0\text{M}} = \frac{1}{2} \int\frac{d^3k}{(2\pi)^3}\sqrt{k_x^2+k_y^2+k_z^2}.
\end{equation}
By adding the correspondent damping function the result yields
\begin{eqnarray}\label{MC}
    \mathcal{E}_{0\text{M}}(\delta) &=&\frac{1}{2} \int\frac{d^3k}{(2\pi)^3}\sqrt{k_x^2+k_y^2+k_z^2}e^{-\delta\sqrt{k_x^2+k_y^2+k_z^2}}\nonumber\\
    &=&\frac{3}{2\pi^2\delta^4},
\end{eqnarray}
which matches exactly the divergent term present in Eq. \eqref{appendixEnDen}. The formal procedure to obtain the renormalized vacuum energy density is performed as follows:
\begin{equation}
    \mathcal{E}_{\text{ren}}^0 =\lim_{\delta\rightarrow 0}\left[\mathcal{E}_0(\delta) - \mathcal{E}_{0\text{M}}(\delta) \right].
\end{equation}
Of course, by using Eqs. \eqref{appendixEnDen} and \eqref{MC} in the above expression we correctly obtain the renormalized vacuum energy density
\begin{equation}
    \mathcal{E}_{\text{ren}}^0 = -\frac{\pi^2}{90d^4},
\end{equation}
which is exactly the same as the one obtained by using the zeta function method in Eq. \eqref{LZEMassless1}. This shows that at least when compared to the cutoff method the vacuum energy density does not show any dependency on the regularization method adopted, as it does in Refs. \cite{Cavalcanti:2003tw, Edery:2005bx}.


\begin{thebibliography}{10}
\bibitem{Casimir1948dh}
H.~B.~G. Casimir, {\it {On the Attraction Between Two Perfectly Conducting
  Plates}},  {\em Indag. Math.} {\bf 10} (1948) 261--263.
\bibitem{Sparnaay1958}
M.~J. Sparnaay, {\it {Measurements of attractive forces between flat plates}},
  {\em Physica} {\bf 24} (1958) 751--764.
\bibitem{PhysRevLett.81.5475}
S.~K. Lamoreaux, {\it Erratum: Demonstration of the casimir force in the 0.6 to
  6 $\mathit{\ensuremath{\mu}}m$ range [phys. rev. lett. 78, 5 (1997)]},  {\em
  Phys. Rev. Lett.} {\bf 81} (Dec, 1998) 5475--5476.
\bibitem{Lamoreaux1996wh}
S.~K. Lamoreaux, {\it {Demonstration of the Casimir force in the 0.6 to 6
  micrometers range}},  {\em Phys. Rev. Lett.} {\bf 78} (1997) 5--8. [Erratum:
  Phys.Rev.Lett. 81, 5475--5476 (1998)].
\bibitem{MohideenRoy1998iz}
U.~Mohideen and A.~Roy, {\it {Precision measurement of the Casimir force from
  0.1 to 0.9 micrometers}},  {\em Phys. Rev. Lett.} {\bf 81} (1998) 4549--4552,
  [\href{https://arxiv.org/abs/physics/9805038}{{\tt physics/9805038}}].
\bibitem{Bressi:2002fr}
G.~Bressi, G.~Carugno, R.~Onofrio, and G.~Ruoso, {\it {Measurement of the
  Casimir force between parallel metallic surfaces}},  {\em Phys. Rev. Lett.}
  {\bf 88} (2002) 041804, [\href{https://arxiv.org/abs/quant-ph/0203002}{{\tt
  quant-ph/0203002}}].
\bibitem{MOSTEPANENKO2000}
V.~M. Mostepanenko, {\it {New experimental results on the Casimir effect}},
  {\em {Brazilian Journal of Physics}} {\bf 30} (06, 2000) 309 -- 315.
\bibitem{PhysRevA.78.020101}
W.~J. Kim, M.~Brown-Hayes, D.~A.~R. Dalvit, J.~H. Brownell, and R.~Onofrio,
  {\it Anomalies in electrostatic calibrations for the measurement of the
  casimir force in a sphere-plane geometry},  {\em Phys. Rev. A} {\bf 78} (Aug,
  2008) 020101, [\href{https://arxiv.org/abs/0812.0028}{{\tt
  arXiv:0812.0028}}].
\bibitem{PhysRevA.81.052115}
Q.~Wei, D.~A.~R. Dalvit, F.~C. Lombardo, F.~D. Mazzitelli, and R.~Onofrio, {\it
  Results from electrostatic calibrations for measuring the casimir force in
  the cylinder-plane geometry},  {\em Phys. Rev. A} {\bf 81} (May, 2010)
  052115, [\href{https://arxiv.org/abs/1101.1476}{{\tt arXiv:1101.1476}}].
\bibitem{BordagMohideenMostepanenko}
M.~Bordag, U.~Mohideen, and V.~M. Mostepanenko, {\it {New developments in the
  Casimir effect}},  {\em Phys. Rept.} {\bf 353} (2001) 1--205,
  [\href{https://arxiv.org/abs/quant-ph/0106045}{{\tt quant-ph/0106045}}].
\bibitem{Mostepanenko:1997sw}
V.~M. Mostepanenko and N.~N. Trunov, {\it {The Casimir effect and its
  applications}},  {\em Oxford, UK: Clarendon (1997) 199 p} (1997).
\bibitem{bordag2009advances}
M.~Bordag, G.~L. Klimchitskaya, U.~Mohideen, and V.~M. Mostepanenko, {\em
  Advances in the Casimir effect}, vol.~145.
\newblock OUP Oxford, 2009.
\bibitem{Milton:2001yy}
K.~A. Milton, {\it {The Casimir effect: Physical manifestations of zero-point
  energy}},  {\em River Edge, USA: World Scientific (2001) 301 p} (2001).
\bibitem{Edery:2006td}
A.~Edery, {\it {Casimir piston for massless scalar fields in three
  dimensions}},  {\em Phys. Rev. D} {\bf 75} (2007) 105012,
  [\href{https://arxiv.org/abs/hep-th/0610173}{{\tt hep-th/0610173}}].
\bibitem{Edery:2005bx}
A.~Edery, {\it {Multidimensional cut-off technique, odd-dimensional Epstein
  zeta functions and Casimir energy of massless scalar fields}},  {\em J. Phys.
  A} {\bf 39} (2006) 685--712,
  [\href{https://arxiv.org/abs/math-ph/0510056}{{\tt math-ph/0510056}}].
\bibitem{Edery:2005aj}
A.~Edery, {\it {Casimir forces in Bose-Einstein condensates: Finite size
  effects in three-dimensional rectangular cavities}},  {\em J. Stat. Mech.}
  {\bf 0606} (2006) P06007, [\href{https://arxiv.org/abs/hep-th/0510238}{{\tt
  hep-th/0510238}}].
\bibitem{Edery:2007xf}
A.~Edery and I.~MacDonald, {\it {Cancellation of nonrenormalizable hypersurface
  divergences and the d-dimensional Casimir piston}},  {\em JHEP} {\bf 09}
  (2007) 005, [\href{https://arxiv.org/abs/0708.0392}{{\tt
  arXiv:0708.0392}}].
\bibitem{Hawking1977}
S.~W. Hawking, {\it Zeta function regularization of path integrals in curved
  spacetime},  {\em Commun. math. Phys.} {\bf 55} (1977) 133--148.
\bibitem{Elizalde2012zv}
E.~Elizalde, {\it {Zeta function regularization in Casimir effect calculations
  and J.S. Dowker's contribution}},  {\em Int. J. Mod. Phys. A} {\bf 27} (2012)
  1260005, [\href{https://arxiv.org/abs/1205.7032}{{\tt arXiv:1205.7032}}].
\bibitem{Lin:2014lva}
Rui-Hui Lin, Xiang-Hua Zhai, {\it {Thermal Casimir effect for rectangular cavities
  inside (D+1)-dimensional Minkowski space-time revisited}},  {\em Int. J. Mod.
  Phys. A} {\bf 29} (2014) 1450043,
  [\href{https://arxiv.org/abs/1402.3924}{{\tt arXiv:1402.3924}}].
\bibitem{Elizalde1994book}
E.~Elizalde, S.~D. Odintsov, A.~Romeo, A.~A. Bytsenko, and S.~Zerbini, {\em
  {Zeta regularization techniques with applications}}.
\newblock World Scientific, 1994.
\bibitem{DowkerCritchley1975tf}
J.~S. Dowker and R.~Critchley, {\it {Effective Lagrangian and Energy Momentum
  Tensor in de Sitter Space}},  {\em Phys. Rev. D} {\bf 13} (1976) 3224.
\bibitem{brandonjicRedBloodCells10}
K.~Bradonjic, J.~Swain, A.~Widom, and Y.~Srivastava, {\it The casimir effect
  in biology: The role of molecular quantum electrodynamics in linear
  aggregations of red blood cells},  {\em Journal of Physics: Conference
  Series} {\bf 161} (05, 2009) 012035.
\bibitem{geometricpot_wettability76}
Peng Liu and Ji-Huan He, {\it Geometric potential: An explanation of nanofiber?s
  wettability},  {\em Thermal Science} {\bf 22} (01, 2017) 146--146.
\bibitem{CasimirUniversalCellMembrane4}
P.~Pawlowski and P.~Zielenkiewicz, {\it The quantum casimir effect may be a
  universal force organizing the bilayer structure of the cell membrane},  {\em
  The Journal of membrane biology} {\bf 246} (04, 2013).
\bibitem{Gambassi2009REV}
A.~Gambassi, {\it {The Casimir effect: From quantum to critical fluctuations}},
   {\em J. Phys. Conf. Ser.} {\bf 161} (2009) 012037,
  [\href{https://arxiv.org/abs/0812.0935}{{\tt arXiv:0812.0935}}].
\bibitem{machta113}
B.~B. Machta, S.~L. Veatch, and J.~P. Sethna, {\it Critical casimir forces in
  cellular membranes},  {\em Phys. Rev. Lett.} {\bf 109} (Sep, 2012) 138101.
\bibitem{QuasiPNanotubes}
Chao-Jun Feng, Xin-Zhou Li, Xiang-Hua Zhai, {\it {Casimir Effect under Quasi-Periodic
  Boundary Condition Inspired by Nanotubes}},  {\em Mod. Phys. Lett. A} {\bf
  29} (2014) 1450004, [\href{http://xxx.lanl.gov/abs/1312.1790}{{\tt
  arXiv:1312.1790}}].
\bibitem{klecioQuasiPNanotubes}
K.~E.~L. de~Farias and H.~F. Santana~Mota, {\it {Quantum vacuum fluctuation
  effects in a quasi-periodically identified conical spacetime}},  {\em Phys.
  Lett. B} {\bf 807} (2020) 135612,
  [\href{https://arxiv.org/abs/2005.03815}{{\tt arXiv:2005.03815}}].
\bibitem{QuantumSpringFromCasimir}
Chao-Jun Feng and Xin-zhou Li, {\it {Quantum Spring from the Casimir Effect}},  {\em
  Phys. Lett. B} {\bf 691} (2010) 167--172,
  [\href{https://arxiv.org/abs/1007.2026}{{\tt arXiv:1007.2026}}].
\bibitem{QuantumSpring}
Chao-Jun Feng and Xin-zhou Li, {\it {Quantum Spring}},  {\em Int. J. Mod. Phys. Conf.
  Ser.} {\bf 7} (2012) 165--173, [\href{https://arxiv.org/abs/1205.4475}{{\tt
  arXiv:1205.4475}}].
\bibitem{QuantumSpringD+1}
Xiang-hua Zhai, Xin-zhou Li, Chao-Jun Feng, {\it {The Casimir force of Quantum Spring
  in the (D+1)-dimensional spacetime}},  {\em Mod. Phys. Lett. A} {\bf 26}
  (2011) 669--679, [\href{https://arxiv.org/abs/1008.3020}{{\tt
  arXiv:1008.3020}}].
\bibitem{GibbonsHaking1976}
G.~W. Gibbons and S.~W. Hawking, {\it {Action Integrals and Partition Functions
  in Quantum Gravity}},  {\em Phys. Rev. D} {\bf 15} (1977) 2752--2756.
\bibitem{bellac1991quantum}
M.~Bellac and G.~Barton, {\em Quantum and Statistical Field Theory}.
\newblock Oxford Science Publ. Clarendon Press, 1991.
\bibitem{Zinn-Justin:572813}
J.~Zinn-Justin, {\em {Quantum Field Theory and Critical Phenomena; 4th ed.}}
\newblock International series of monographs on physics. Clarendon Press,
  Oxford, 2002.
\bibitem{PhysRevD.83.104042}
V.~B. Bezerra, G.~L. Klimchitskaya, V.~M. Mostepanenko, and C.~Romero, {\it
  Thermal casimir effect in closed friedmann universe revisited},  {\em Phys.
  Rev. D} {\bf 83} (May, 2011) 104042.
\bibitem{Bezerra:2011nc}
V.~B. Bezerra, V.~M. Mostepanenko, H.~F. Mota, and C.~Romero, {\it {Thermal
  Casimir effect for neutrino and electromagnetic fields in closed Friedmann
  cosmological model}},  {\em Phys. Rev. D} {\bf 84} (2011) 104025,
  [\href{https://arxiv.org/abs/1110.4504}{{\tt arXiv:1110.4504}}].
\bibitem{Bezerra:2017zqq}
V.~B. Bezerra, H.~F. Mota, and C.~R. Muniz, {\it {Casimir Effect in the Rainbow
  Einstein's Universe}},  {\em EPL} {\bf 120} (2017), no.~1 10005,
  [\href{https://arxiv.org/abs/1708.02627}{{\tt arXiv:1708.02627}}].
\bibitem{mellin}
Y.~A. Brychkov, O.~I. Marichev, and N.~V. Savischenko, {\em Handbook Mellin
  Transforms}.
\newblock Advances in Applied Mathematics. CRC Press, 2019.
\bibitem{BezerradeMello2011nv}
E.~R. Bezerra~de Mello and A.~A. Saharian, {\it {Topological Casimir effect in
  compactified cosmic string spacetime}},  {\em Class. Quant. Grav.} {\bf 29}
  (2012) 035006, [\href{https://arxiv.org/abs/1107.2557}{{\tt
  arXiv:1107.2557}}].
\bibitem{MotaDispiration}
H.~F. Mota, E.~R. Bezerra~de Mello, and K.~Bakke, {\it {Scalar Casimir effect
  in a high-dimensional cosmic dispiration spacetime}},  {\em Int. J. Mod.
  Phys. D} {\bf 27} (2018), no.~12 1850107,
  [\href{https://arxiv.org/abs/1704.01860}{{\tt arXiv:1704.01860}}].
\bibitem{sonego2010}
S.Sonego~, {\it {Ultrastatic spacetimes}}, {\em J. Math. Phys.} {\bf 51} (2010), 092502, [\href{https://arxiv.org/abs/1004.1714}{{\tt arXiv:1004.1714}}].
\bibitem{PontualMoraes}
I.~Pontual and F.~Moraes, {\it Casimir effect around a screw dislocation},
  {\em Philosophical Magazine A} {\bf 78(5)} (1998) 1073--1084.
\bibitem{Cognola:1993qg}
G.~Cognola, K.~Kirsten, and L.~Vanzo, {\it {Free and selfinteracting scalar
  fields in the presence of conical singularities}},  {\em Phys. Rev. D} {\bf
  49} (1994) 1029--1038, [\href{https://arxiv.org/abs/hep-th/9308106}{{\tt
  hep-th/9308106}}].
\bibitem{Bytsenko:1992hh}
A.~A. Bytsenko, G.~Cognola, and L.~Vanzo, {\it {Vacuum energy for
  (3+1)-dimensional space-time with compact hyperbolic spatial part}},  {\em J.
  Math. Phys.} {\bf 33} (1992) 3108--3111. [Erratum: J.Math.Phys. 34, 1614
  (1993)].
\bibitem{Landau}
L.~D. Landau and E.~M. Lifshitz, {\em Statistical Physics, Part I}.
\newblock Pergamon Press, Oxford, 1980.
\bibitem{abramowitz}
M.~Abramowitz and I.~Stegun, {\em Handbook of mathematical functions}.
\newblock National Bureau of Standards, Washington U.S.A, 1964.
\bibitem{gradshteyn2000table}
I.~S. Gradshtein and I.~M. Ryzhik, {\em Table of integrals, series, and
  products}.
\newblock Academic Press, 1980.
\bibitem{Elizalde1994}
E.~Elizalde, {\it Analysis of an inhomogeneous generalized epstein-hurwitz
  zeta function with physical applications},  {\em Journal of Mathematical
  Physics} {\bf 35} (1994) 6100--6122.
\bibitem{LandauStatistical}
L.~D. Landau and E.~M. Lifshitz, {\em Statistical Physics, Part I}.
\newblock Pergamon Press, Oxford, 1980.
\bibitem{sinden1994dna}
R.~Sinden, {\em DNA Structure and Function}.
\newblock Elsevier Science, 1994.
\bibitem{RNASize}
A.~M. Yoffe, P.~Prinsen, A.~Gopal, C.~M. Knobler, W.~M. Gelbart, and
  A.~Ben-Shaul, {\it Predicting the sizes of large rna molecules},  {\em
  Proceedings of the National Academy of Sciences} {\bf 105} (2008), no.~42
  16153--16158,
  [\href{https://www.pnas.org/content/105/42/16153.full.pdf}{{\tt
  PNAS/105/42/16153}}].
\bibitem{RNASmallSize}
G.~Storz, {\it An expanding universe of noncoding rnas},  {\em Science} {\bf
  296} (2002), no.~5571 1260--1263.
\bibitem{Elizalde:2020mlv}
E.~Elizalde, {\it {Zeta Functions and the Cosmos\textemdash{}A Basic Brief
  Review}},  {\em Universe} {\bf 7} (2020), no.~1 5.
\bibitem{Cavalcanti:2003tw}
R.~M. Cavalcanti, {\it {Casimir force on a piston}},  {\em Phys. Rev. D} {\bf
  69} (2004) 065015, [\href{https://arxiv.org/abs/quant-ph/0310184}{{\tt
  quant-ph/0310184}}].

\end{thebibliography}
\bibliographystyle{JHEP}
\end{document}